\documentclass[authoryear]{elsarticle}

\usepackage{rotating}
\usepackage{lscape}
\usepackage{graphicx,epsfig,epsf,graphics,epstopdf}
\usepackage{hyperref}
\usepackage{mhchem,hhline}
\usepackage[normalem]{ulem}
\usepackage{caption}
\usepackage{float}
\usepackage{natbib}

\def\apj{ApJ}
\def\nat{Nature}
\def\mnras{MNRAS}
\def\apjl{ApJL}
\def\jgr{JGR}
\def\apjs{ApJS}
\def\araa{ARAA}

\pdfoutput=1 
\usepackage{amsmath,amstext,amssymb}
\usepackage[T1]{fontenc}
\usepackage{apjfonts}
\usepackage{tikz}
\usetikzlibrary{arrows}
\usetikzlibrary{positioning}
\usetikzlibrary{shapes,snakes}
\usepackage[figure,figure*]{hypcap}

\makeatletter
\let\saved@includegraphics\includegraphics
\AtBeginDocument{\let\includegraphics\saved@includegraphics}
\renewenvironment*{figure}{\@float{figure}}{\end@float}
\makeatother

\begin{document}
\begin{frontmatter}
\title{Hydrogen Cyanide in Nitrogen-Rich Atmospheres of Rocky Exoplanets}

\author[add1]{P.~B. Rimmer}
\ead{pbr27@cam.ac.uk}
\author[add2,add3]{S. Rugheimer}
\ead{srm26@st-andrews.ac.uk}
\address[add1]{University of Cambridge Department of Earth Sciences, Downing St, Cambridge CB2 3EQ and University of Cambridge, Cavendish Astrophysics, JJ Thomson Ave, Cambridge CB3 0HE and MRC Laboratory of Molecular Biology, Francis Crick Ave, Cambridge CB2 OQH}
\address[add2]{Centre for Exoplanet Science, University of St. Andrews, School of Earth and Environmental Sciences, Irvine Building, North Street, St. Andrews, KY16 9AL, UK}
\address[add3]{University Oxford, Atmospheric, Oceanic, and Planetary Physics Department, Clarendon Laboratory, Sherrington Road, Oxford, OX1 3PU}

\begin{abstract}
\setcounter{table}{0}
\renewcommand{\thetable}{5.\arabic{table}}
Hydrogen cyanide (HCN) is a key feedstock molecule for the production of life's building blocks. The formation of HCN in an \ce{N_2}-rich atmospheres requires first that the triple bond between N$\equiv$N be severed, and then that the atomic nitrogen find a carbon atom. These two tasks can be accomplished via photochemistry, lightning, impacts, or volcanism. The key requirements for producing appreciable amounts of HCN are the free availability of \ce{N_2} and a local carbon to oxygen ratio of C/O $\geq 1$. We discuss the chemical mechanisms by which HCN can be formed and destroyed on rocky exoplanets with Earth-like \ce{N_2} content and surface water inventories, varying the oxidation state of the dominant carbon-containing atmospheric species. HCN is most readily produced in an atmosphere rich in methane (\ce{CH_4}) or acetylene (\ce{C_2H_2}), but can also be produced in significant amounts ($> 1$ ppm) within CO-dominated atmospheres. Methane is not necessary for the production of HCN. We show how destruction of HCN in a \ce{CO_2}-rich atmosphere depends critically on the poorly-constrained energetic barrier for the reaction of HCN with atomic oxygen. We discuss the implications of our results for detecting photochemically produced HCN, for concentrating HCN on the planet's surface, and its importance for prebiotic chemistry.
\end{abstract}
\end{frontmatter}

\renewcommand{\arraystretch}{0.6}


\section{Introduction}

Hydrogen cyanide (HCN) is a key molecule for prebiotic photochemistry. HCN is a precursor molecule to amino acids via Strecker synthesis, where HCN is reacted with ammonia and an aldehyde to form various amino acids \citep{Strecker1854}, and has for this reason been invoked as the mechanism for generating the amino acid glycine in the Miller-Urey experiment \citep{Miller1957}. The structure of the nucleobase adenine (\ce{C_5H_5N_5}) is effectively five connected HCN molecules. Adenine can be formed by HCN reacting with its conjugate base and polymerizing, followed by photochemical rearrangement of the resulting HCN tetramer and further HCN polymerization \citep{ferris1966,sanchez1966,Sanchez1967,sanchez1970}.

More recently, HCN has been discovered to be present at each of the initial photochemical reactions that produce lipids, amino acids and nucleosides, the three building blocks of life \citep{ritson2012,patel2015,Xu2018}. In this scenario, UV light is absorbed by hydrogen cyanide and hydrogen sulfide or bisulfite in water. The UV photon initiates photodetachment of an electron from the hydrogen sulfide or bisulfite, and the solvated electron reduces the hydrogen cyanide, forming an imine complex that, when hydrolised, forms formaldehyde. This photochemical homologation can continue, with the dissociated products of cyanide reacting with the product sugars. 

Other plausible feedstock molecules generated from reactions involving HCN, such as cyanamide and cyanoacetylene, react with the sugars to form the pyrimidine nucleotides. In addition, these feedstock molecules, along with methylamine and nitrogen oxides, act as activating agents to facilitate the ligation of these building blocks into polypeptides, a key next step in prebiotic chemistry \citep{Mariani2018}.

HCN can be produced photochemically \citep{Zahnle1986,Tian2011}, by interaction of the atmospheric gas with energetic particles \citep{Air2016}, by lightning \citep{Chameides1981,Ardaseva2017}, and by meteor impacts \citep{Ferus2017}. In each of these scenarios, the authors have recognized that reduced carbon-bearing species greatly facilitate the production of hydrogen cyanide. In some circumstances, especially with impacts, up to 1\% of the affected gas can be converted into HCN, but again only with a relatively reduced gas, and in the presence of significant concentrations of molecular nitrogen or ammonia. The same events occurring in a gas of different composition, however, can result in trace amounts of HCN, less than 1 ppb. 

\begin{figure}
\includegraphics[width=\textwidth]{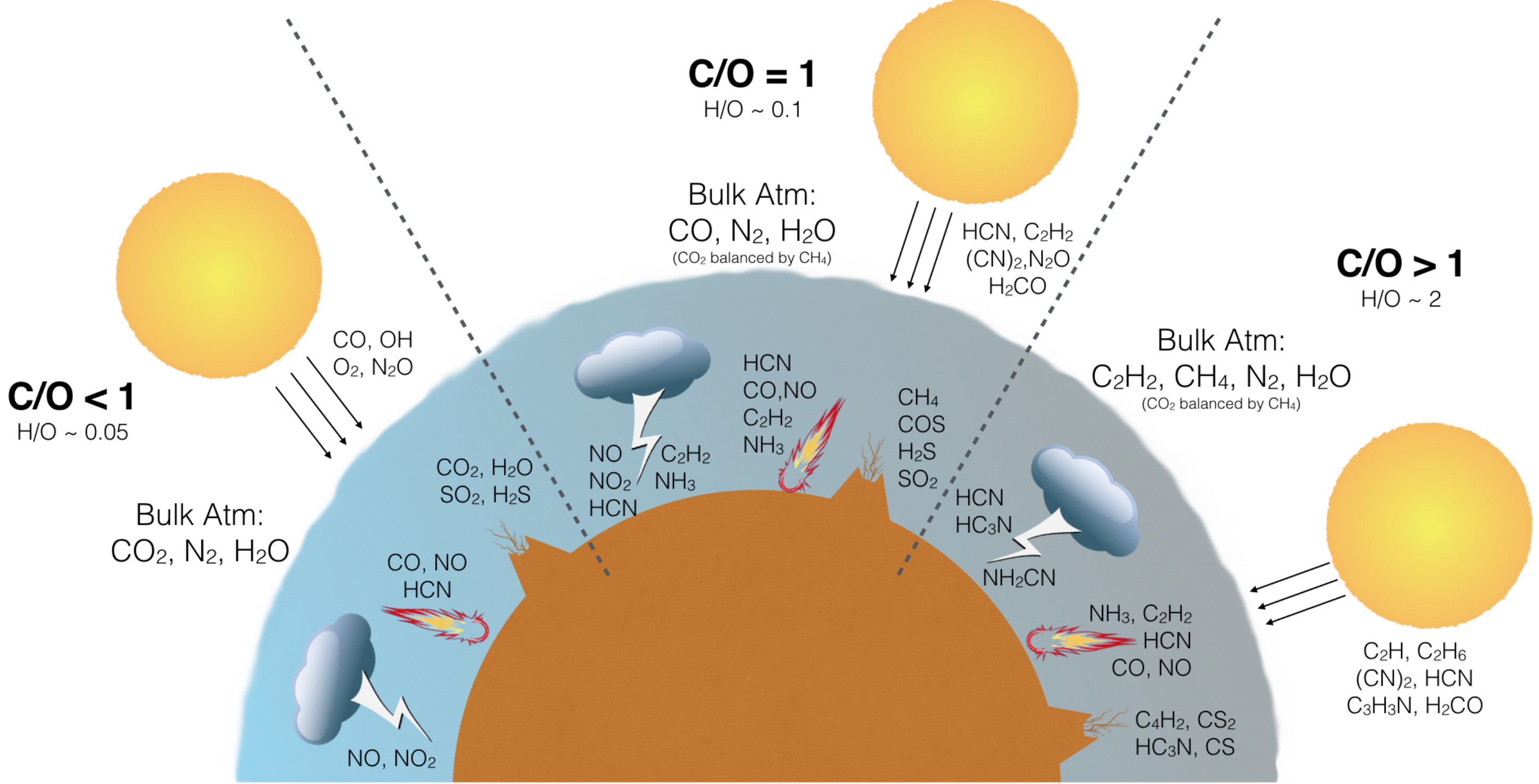}
\caption{A representation of how the bulk chemistry changes with the bulk C/O ratio, and how more-or-less local chemical processes: lightning, impacts, volcanism, ultraviolet photons, are affected by the C/O ratio, from C/O < 1 to C/O > 1. \label{fig:hcn-cartoon}}
\end{figure}

Exoplanets can be seen as diverse laboratories for atmospheric and prebiotic chemistry, and so we need not be bound by constraints from early Earth atmosphere. Bulk atmospheric composition is given for rocky planet atmospheres under a wide variety of elemental abundance ratios \citep{Hu2014}. In addition, changes in the water inventory \citep{Gao2015}, the amount of hydrogen retained in the atmosphere \citep{Wordsworth2011}, and the concentration of molecular nitrogen \citep{Wordsworth2014}, can have profound effects on the evolution of the planet's atmosphere. Here we will focus on more Earth-like planets, in terms of their nitrogen content and water inventories. Given the presence of molecular nitrogen as a dominant atmospheric constituent, the bulk atmospheric composition will here be represented largely in terms of the C/O ratio (see Fig. \ref{fig:hcn-cartoon}).

We present our methods for calculating the climate and photochemistry in Section \ref{sec:methods}, and then discuss the connection between the C/O ratio, the relative UV flux and HCN in Section \ref{sec:theory}. In Section \ref{sec:results} present the concentrations of HCN as a function of the C/O ratio, UV flux, and the barrier to destruction by atomic oxygen. We conclude with a discussion of these results (Section \ref{sec:conclusion}).

\section{Methods}
\label{sec:methods}

\begin{figure}
\includegraphics[width=\textwidth]{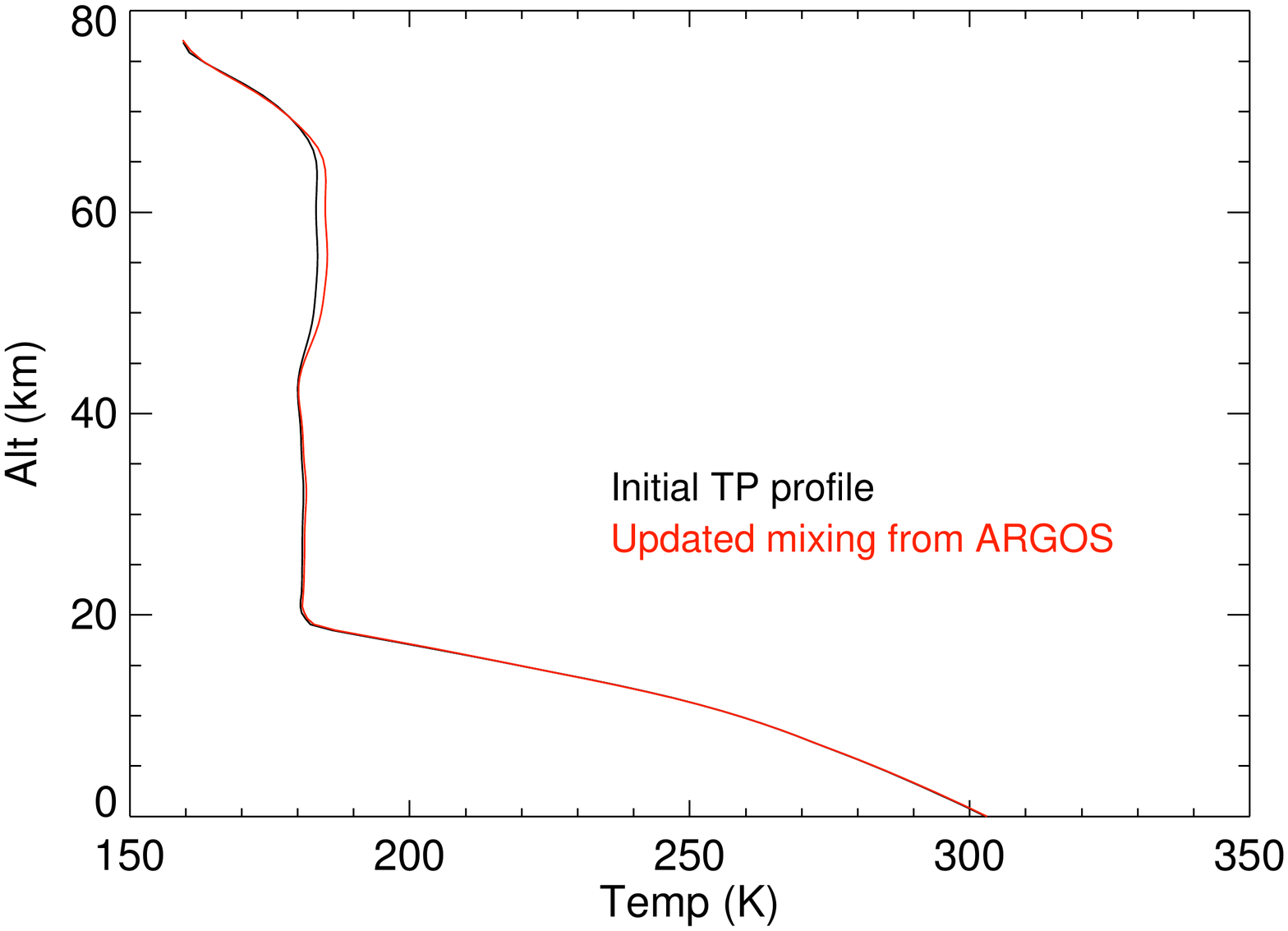}
\caption{The temperature [K] as a function of altitude [km] for an anoxic 1 bar surface pressure atmosphere composed of 89\% N$_2$, 10\% CO$_2$ and 1\% CH$_4$ (black line), and with updated mixing ratios taken from the ARGOS photochemical model using the STAND2019 chemical network (red line).}\label{fig:t-profile}
\end{figure}

To calculate the temperature and pressure profiles, we use a 1D climate model developed for high-CO$_2$/high-CH$_4$ terrestrial atmospheres \citep{pavlov2000,kharecha2005,haqq2008}. This model simulates an anoxic 1 bar atmosphere composed of 89\% N$_2$, 10\% CO$_2$ and 1\% CH$_4$, similar to some of the models of \citet{Zahnle1986}.Gases apart from H$_2$O are assumed to be well mixed in the atmosphere. The atmosphere was divided into 101 layers. The code uses a $\delta$ two-stream scattering approximation for the absorption of solar radiation \citep{Toon1989} and 4-term, correlated \textit{k}-coefficients to parameterize the absorption by the main greenhouse gases, namely: H$_2$O, CH$_4$, and CO$_2$.  We use the solar evolution model for the incoming stellar radiation at 3.9 Ga \citep{claire2012}, both for the climate modelling and the UV spectrum for the photochemistry discussed below. The temperature profile from the 1D climate model is in Fig. \ref{fig:t-profile}.

We take the prescribed stellar spectra and surface mixing ratios, and the temperature profile determined by the climate model, and apply them to the ARGO photochemistry/diffusion model \citep{Rimmer2016}, which solves the set of photochemical kinetics equations for a moving parcel of atmospheric gas:
\begin{equation}
\dfrac{\partial n_i}{\partial t} = P_i - L_i - \dfrac{\partial \Phi_i}{\partial z},
\end{equation}
Here, $n_i$ [cm$^{-3}$] is the number density of species $i$, where $i = 1,...,I_s$, $I_s$ being the total number of species. The quantity $\Phi_i$ is the flux of the species $i$ into and out of the parcel when it is at height $z$. The parcel is used to give a chemical profile of the atmosphere, and UV photo-transport is calculated in order to determine the efficiency of the photochemistry at each parcel height \citep{Rimmer2016}. $P_i$ [cm$^{-3}$ s$^{-1}$] and $L_i$ [cm$^{-3}$ s$^{-1}$] represent the production and loss rates, which are determined by using the a modified version of the STAND2016 chemical network \citep{Rimmer2016}, which we hereafter will refer to as STAND2019. The network is described in detail below. This model will determine a different chemical profile from the climate model, but does not update the temperature profile. The deviation between the temperature profile for the initial and updated mixing ratios is small, and is shown in Fig. \ref{fig:t-profile}.

The STAND2019 chemical network is a modification of the STAND2016 network, which includes a robust H/C/N/O chemistry with species of up to 3 carbons, 2 nitrogens, 3 oxygens, and 8 hydrogens, valid for temperatures up to 30000 K, and including some relevant gas-phase reactions involving Na, K, Mg, Fe, Ti, Si, and Cl. Altogether, the STAND2019 chemical network comprises over 5000 reactions involving over 350 chemical species. The full network is included in the Supplementary Information. We now describe the most significant differences between STAND2016 and STAND2019. The reverse of a reaction relevant for formaldehyde in STAND2016 has been suppressed in the STAND2019 network in order to properly reproduce the formaldehyde (HCHO) abundances observed in Earth's atmosphere today \citep{Granville2018}.  The relevant reaction is:
\begin{equation}
\ce{H} + \ce{CO} + \ce{M} \rightarrow \ce{HCO} + \ce{M}.
\label{eqn:forward}
\end{equation}
and its reverse is:
\begin{equation}
\ce{HCO} + \ce{M} \rightarrow \ce{H} + \ce{CO} + \ce{M}.
\label{eqn:rev}
\end{equation}
The reverse rate is calculated using the Gibbs free energy, as described by \cite{Rimmer2016}. This is so that the chemical steady state converges to thermochemical equilibrium at sufficiently long timescales in the absence of irreversible reactions (e.g. photochemistry). This is important for reproducing hot exoplanet atmospheric compositions and the gas-phase component of magma chemistries, both applications for the STAND2019 network. The consequence of this approach is that sometimes the barrier calculated by reversing the reaction does not reproduce the actual chemical barrier to the reaction, and in the case of R\ref{eqn:rev}, the barrier is greatly underestimated. In order to compensate for this, the rate constant for R3 is set to: 
\begin{equation*}
k = 6.0 \times 10^{-11} \, e^{-5370 \, K / T} \, \text{cm$^{3}$ s$^{-1}$},
\end{equation*}
and using this rate constant, with this sizable barrier, reproduces the the formaldehyde (HCHO) abundances \citep{Granville2018}.

The forward and reverse barriers for this reaction will need to be revisited in the future to figure out why the barrier for R\ref{eqn:rev} is considerably higher than determined when reversing R\ref{eqn:forward}.

In addition, the relevant reaction from STAND2016 has been replaced with R\ref{eqn:HCN-dest}:
\begin{equation*}
\ce{HCN} + \ce{O} \rightarrow \ce{NCO} + \ce{H},
\end{equation*}
with the rate constant of:
\begin{equation}
k = 1.43 \times 10^{-12} \; \Bigg(\dfrac{T}{300 \, {\rm K}} \Bigg)^{\!\!1.47} \, e^{-3770 \, K / T} \, \text{cm$^{3}$ s$^{-1}$},
\end{equation}
from \citet{perry1985}. This choice of rate constant will be justified in Section \ref{sec:atomic-o}. Finally, we have added the reaction \citep{Becker1983}:
\begin{equation}
\ce{HCN} + \ce{C_2H} \rightarrow \ce{HC_3N} + \ce{H},
\end{equation}
for which we assign a rate constant of:
\begin{equation}
k = 2.2 \times 10^{-12} \, \text{cm$^{3}$ s$^{-1}$}.
\end{equation}
This reaction was added to see if cyanoacetylene might be generated in trace but significant quantities in our models, which would be relevant for prebiotic chemistry, as acyanoacetylene is one of the key feedstock molecules for forming some of the canonical amino acids and the pyrimidine nucleotides. It turns out that no significant quantities of cyanoacetylene are generated in any of the atmospheric models we consider.

Changes to the methanol chemistry \citep{Tsai2017}, are included, and not expected to make much difference here. The STAND2019 network, which appears first in this paper with new reaction numbers, is included in full in the Supplementary Information. The number of reactions is large (over 5000), but only a relatively small number of reactions are relevant for the HCN chemistry. The reactions relevant for atmospheric HCN production, by UV photons as well as by impacts, lighting and energetic particle interactions, are discussed in detail in Section \ref{sec:theory}. These reactions depend on the abundances of other species, which are determined by yet other reactions, and so we use the entire network when modeling the atmospheric photochemistry.

\begin{figure}
\includegraphics[width=\textwidth]{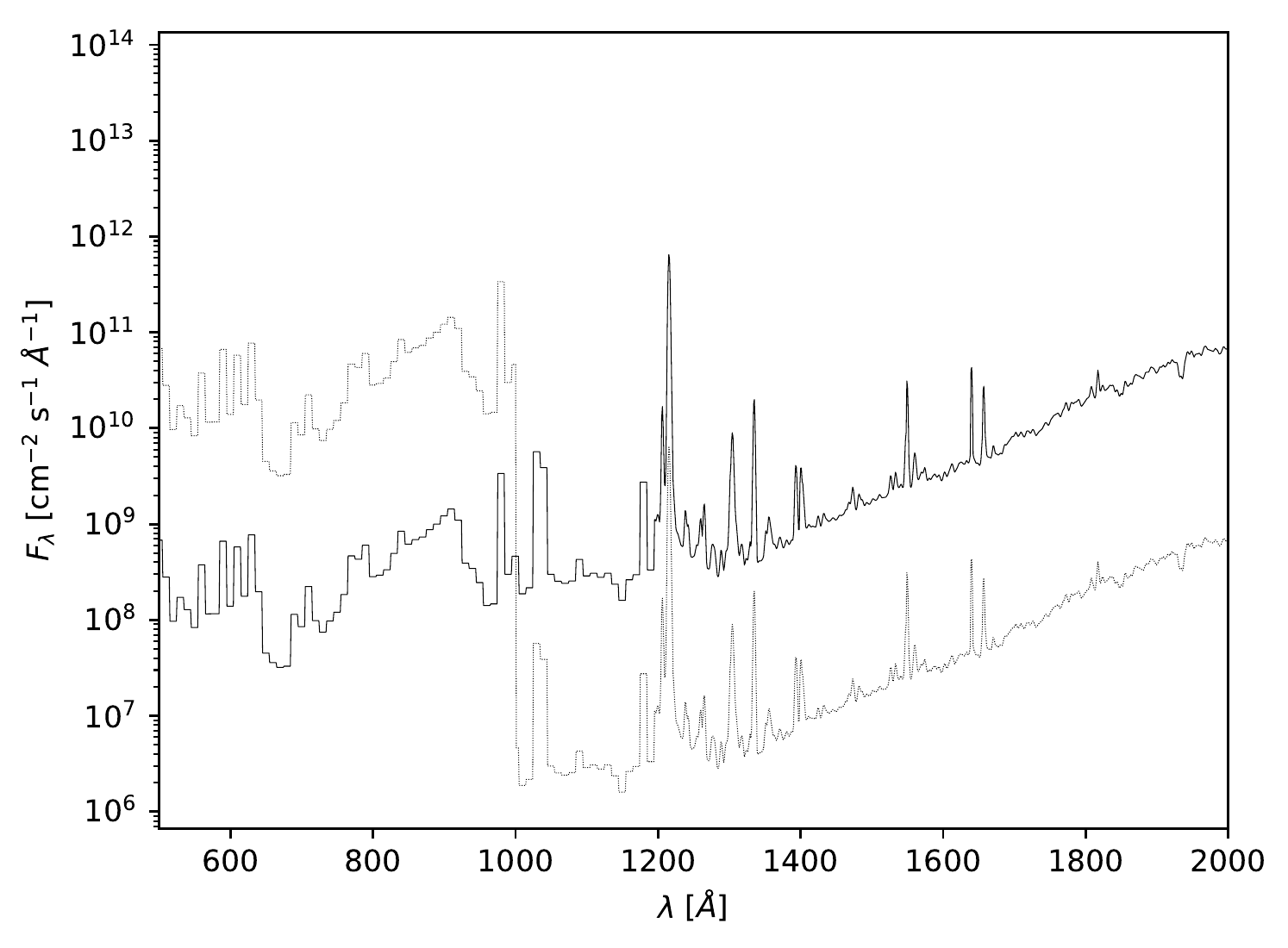}
\caption{The spectrum of the young Sun (solid line) \citep{claire2012}, artifically enhanced between 500 \AA - 1000 \AA\ and suppressed between 1000 \AA - 2000 \AA (dashed line), used to generate the results in Section \ref{sec:co-uv}. \label{fig:alter-spec}}
\end{figure}

Finally, for the alterations in the stellar spectrum in Section \ref{sec:co-uv}, we take a model spectrum \citep{claire2012}, and artifically enhance or reduce the spetrum within the regions of 500 \AA - 1000 \AA\ and 1000 \AA - 2000 \AA, as shown in Fig. \ref{fig:alter-spec}, to test the impact of the UV spectrum on HCN formation.

\section{Theory}
\label{sec:theory}

In an exoplanet atmosphere composed of molecules bearing hydrogen, carbon, oxygen and nitrogen, there are two necessary conditions for producing HCN. First, the nitrogen bonds need to be cleaved, and then second, carbon needs to win over oxygen in the competition for the free nitrogen. The first condition can be relatively easy if the nitrogen is primarily in the form of ammonia (\ce{NH_3}), which has a large absorption cross section over a wider wavelength range than \ce{N_2} ($\lesssim 300$ nm for ammonia compared to $\lesssim 160$ nm for \ce{N_2}). Hot temperatures can also be sufficient to cleave a hydrogen atom from ammonia. Nitrogen in the form of \ce{N_2}, with a strong triple bond, is far more difficult to break apart. Impacts and lightning are sufficient to dissociate \ce{N_2} into N + N, as are UV photons of sufficiently short wavelength ($\lesssim 110$ nm), although these are largely attenuated below the thermosphere by \ce{N_2} itself, assuming that \ce{N_2} is a prevalent atmospheric constituent.

After the nitrogen bonds are broken, the nitrogen atoms react readily with nearby molecules, including some otherwise stable molecules. The simplest (and often the most common) case is if a lone nitrogen atom finds another lone nitrogen atom, in which case it will react to form \ce{N_2}. If there is a great deal of hydrogen available, whether in the form of H or \ce{H_2}, the nitrogen will rapidly react to form NH or \ce{NH_2}, and then will continue to react with the abundant hydrogen to end in ammonia. On the other hand, if the nitrogen atom collides with an oxygen atom, the resulting eventual product will generally be NO or \ce{NO_2}. If the nitrogen atom finds carbon, the eventual product will generally be HCN. Thus the balance of carbon to oxygen in a given geological environment is a key metric to understanding abiotic nitrile chemistry. Fig's \ref{fig:pathway-to-form} and \ref{fig:pathway-to-break} provide the reaction networks for the formation and destruction of HCN under a variety of circumstances.

\begin{figure}[h!]
\centering
\resizebox{.7\linewidth}{!}{
\begin{tikzpicture}[->,>=stealth',shorten >=1pt,auto,node distance=3cm,
                    thick,main node/.style={circle,draw,font=\sffamily\Large\bfseries},hcn node/.style={regular polygon,regular polygon sides=4,draw,font=\sffamily\Large\bfseries},cn node/.style={diamond,draw,font=\sffamily\Large\bfseries}]

  \node[main node] (1) {CH$_4$};
  \node[main node] (2) [right of=1] {CH};
  \node[main node] (3) [right of=2] {CH$_2$};
  \node[main node] (4) [right of=3] {CH$_3$};
  \node[hcn node] (5) [below right=1cm and 1.5cm of 2] {HCN};
  \node[cn node] (6) [below=1cm of 2] {CN};
  \node[main node] (7) [below=1.1cm of 1] {CO};
  \node[main node] (8) [below=1.5cm of 7] {C};
  \node[main node] (9) [right of=8] {NCO};
  \node[main node] (10) [right=1.5cm of 9] {CNN};
  \node[main node] (11) [right=1.5cm of 5] {C$_2$N};
  \node[main node] (12) [left of=7] {CO$_2$};
  \node[main node] (13) [right of=4] {C$_2$H$_6$};

  \path[every node/.style={font=\sffamily\small}]
    (1) edge node [above] {$h\nu$} (2)
    (1) edge [bend left] node [above right=-0.1cm and 0.3cm] {$h\nu$} (3)
    (1) edge [bend left] node [above] {O, OH, or high $T$} (4)
    (2) edge [bend left] node [above left=-0.1cm and 0.5cm] {\ce{H_2}} (4)
    (2) edge node [right] {NO} (5)
    (2) edge node [left] {N$_2$O} (5)
    (3) edge node [right] {N} (5)
    (4) edge node [left] {N} (5)
    (3) edge node [above] {H$_2$} (4)
    (3) edge node [below] {CH$_4$} (4)
    (7) edge node [above] {N} (6)
    (7) edge node [below] {(high T)} (6)
    (7) edge node [right] {$h\nu$} (8)
    (7) edge node [left] {high $T$} (8)
    (7) edge node [left] {$h\nu$+\ce{H_2}} (2)
    (8) edge [bend right] node [below] {N$_2$} (10)
    (7) edge node [right] {NO} (9)
    (9) edge node [left] {H} (5)
    (10) edge node [left] {H} (5)
    (11) edge node [above] {H$_2$O} (5)
    (6) edge node [above] {H,H$_2$} (5)
    (6) edge node [below] {H$_2$O} (5)
    (12) edge node [above] {$h\nu$} (7)
    (12) edge node [below] {high $T$} (7)
    (4) edge node [above] {CH$_3$} (13);
\end{tikzpicture}}
\caption{Collection of known pathways to form HCN on rocky planets. HCN can be produced photochemically (this paper, as well as \citealt{Zahnle1986} and \citealt{Tian2011}), by impacts \citep{Ferus2017}, and by lightning \citep{Ardaseva2017}. The rate constants for these reactions can be found in the STAND2019 Network (Supplementary Information). \label{fig:pathway-to-form}}
\end{figure}
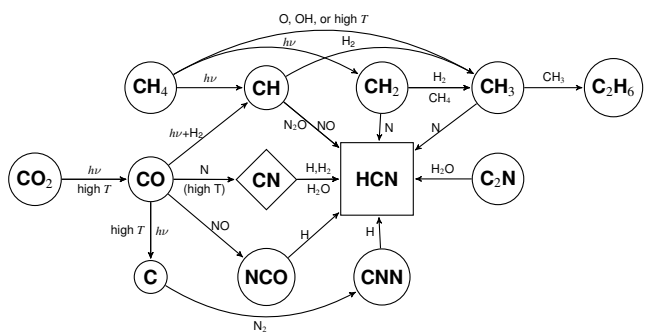

\begin{figure}[h!]
\centering
\resizebox{.6\linewidth}{!}{
\begin{tikzpicture}[->,>=stealth',shorten >=1pt,auto,node distance=3cm,
                    thick,main node/.style={circle,draw,font=\sffamily\Large\bfseries},hcn node/.style={regular polygon,regular polygon sides=4,draw,font=\sffamily\Large\bfseries},cn node/.style={diamond,draw,font=\sffamily\Large\bfseries}]

  \node[hcn node] (1) {HCN};
  \node[cn node] (2) [right of = 1]{CN};
  \node[main node] (3) [left=2cm of 1] {CN$^{-}$};
  \node[main node] (4) [left of = 3] {ACN};
  \node[main node] (5) [right of = 2] {N$_2$};
  \node[main node] (6) [below = 1.5cm of 1] {HNCOH};
  \node[main node] (7) [right of = 6] {NCO};
  \node[main node] (8) [left of = 6] {CO};
  \node[main node] (9) [below = 1.2cm of 6] {HNCO};
  \node[main node] (10) [right of = 7] {NO};

  \path[every node/.style={font=\sffamily\small}]
    (1) edge node [above] {$h\nu$} (2)
    (1) edge node [above] {CH$_3$O$^-$} (3)
    (1) edge node [right] {O} (7)
    (1) edge node [left] {O} (8)
    (3) edge node [above] {A$^+$} (4)
    (2) edge node [above] {N,NO} (5)
    (7) edge node [above] {N} (5)
    (7) edge node [above] {O} (10)
    (1) edge node [left] {OH} (6)
    (6) edge node [left] {OH} (9)
    (9) edge node [left] {$h\nu$} (7);
\end{tikzpicture}}
\caption{Collection of known pathways to destroy HCN on rocky planets. Destruction is primarily via atomic oxygen, the hydroxyl radical, anion chemistry, or photodissociation of HCN followed by reaction with atomic nitrogen or NO. The rate constants for these reactions can be found in the STAND2019 Network (Supplementary Information).
\label{fig:pathway-to-break}}
\end{figure}
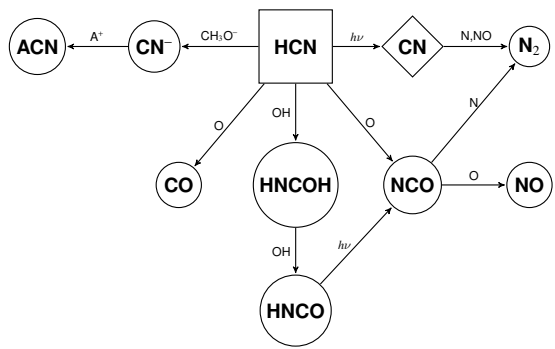

There are two ways to look at this competition: in terms of equilibrium and kinetics. Chameides \& Walker were first to examine the thermochemical equilibrium of NO and HCN. They discuss the competition between carbon and oxygen for the nitrogen \citep{Chameides1981}, and express this competition in terms of the oxidation state of the carbon, where if the available carbon is mostly in the form of carbon monoxide (CO) and methane (\ce{CH_4}), more HCN will result, and if it is mostly in the form of \ce{CO_2}, more NO will result. \citet{Chameides1981} show this relation explicitly in terms of the C/O ratio in their Fig. 3.

We examine the thermochemical formation of HCN as a function of the C/O ratio, and find that we reproduce the results of \citet{Chameides1981}. Where C/O < 1 and the nitrogen bonds are broken, NO is formed in large quantities and HCN in trace quantities. If the C/O > 1 and the nitrogen bonds are cleaved, the opposite holds, and a great deal of HCN is formed with very little NO. If C/O $\approx 1$, then both HCN and NO are formed in reasonable quantities. These results match those of \citet[][their Fig. 3]{Chameides1981}. Methane is not necessary. All of the carbon could be in the form of CO or \ce{C_2H_2}. What is necessary is an elemental C/O ratio $\geq 1$ and a reasonable amount of hydrogen, whether bound as methane, acetylene, molecular hydrogen, or even water vapor. 

\subsection{Mechanisms for Generating and Destroying HCN on Rocky Exoplanets}
\label{sec:mechanism}


The mechanisms by which HCN is formed and destroyed vary with atmospheric composition, temperature, stellar activity, impact rate, surface processes, and lighting rates. The chemical mechanisms themselves do not change for each process, so it is helpful to look at the chemistry as a whole. Fig. \ref{fig:pathway-to-form} illustrates the pathways that can form HCN from \ce{CO_2}, CO, \ce{C_2N} (the most abundant CN-containing species produced by lightning on the Early Earth \citep{Ardaseva2017}), as well as the formation from \ce{CH_4} \citep{Zahnle1986,Tian2011,Air2016}. Some of the reactions are photochemical, denoted by $h\nu$, and require photons of a given range of wavelengths, and these may involve reactive intermediates, e.g. \ce{H_2CN}, in the case of \ce{CH_2 + N}. Other reactions are driven by energetic particles, and some proceed at high temperatures. These driving factors are rarely exclusive. 

On some exoplanets, especially those with high C/O ratios, there may be a significant amount of carbon monoxide present in the atmosphere to begin with. For volcanically degassed atmospheres, the composition will largely depend on the outgassing pressure and redox state of the crust and upper mantle \citep{Gaillard2014}. Otherwise, CO is produced in reasonable quantities by the photodissociation or thermal dissociation of \ce{CO_2} ($< 195$ nm, $E_a \sim 415$ kJ/mol). At very high temperatures ($E_a =$ 300 kJ/mol), CO + N $\rightarrow$ CN + O, and once temperatures drop, the CN can proceed to HCN. CO itself can be dissociated ($< 110$ nm for CO(X$^1\Sigma^+$), $< 210$ nm for CO($A^3\Pi$) , $E_a = 1000$ kJ/mol).

The destruction of HCN (see Fig. \ref{fig:pathway-to-break}) proceeds on both the present and Early Earth primarily via oxidation by atomic oxygen and hydroxyl radicals (e.g. R\ref{eqn:HCN-dest}). In environments like Titan, with very high C/O ratios, where there is virtually no available oxygen, HCN is destroyed by photodissociation into CN followed by electron exchange from anions to form CN$^-$. 

HCN can also react with the hydroxyl radical and a third body in order to form HNCOH, a radical precursor to formamide, potentially relevant to prebiotic chemistry \citep{Saladino2001,Ferus2015}. Most often, though, HNCOH will encounter a second OH radical, forming isocyanic acid (HNCO). Isocyanic acid is quite susceptible to photodissociation, resulting in H + NCO.

Finally, HCN can become absorbed into water droplets, and there has the opportunity to react with a solvated electron, leading to the formation of formaldehyde, or its conjugate base can react with formaldehyde to form glycolonitrile. These are the initial steps toward RNA precursors \citep{ritson2012}. The Henry's Law constant for HCN is about 10 M/bar, and so resulting concentration of HCN in raindrops from the model atmosphere we present blow (C/O ratio of 0.55) is at most 10 nM, which is too small a concentration to be useful for prebiotic chemistry. If the atmospheric C/O ratio approaches 1, the surface mixing ratios could become as high as 10 ppm, and the resulting concentrations at 0.1 mM, on the cusp of prebiotically useful concentrations. The concentrations could be enhanced by overcoming Henry's Law, e.g. by the absorption of formaldehyde (HCHO) by raindrops, and the reaction of formaldehyde with HCN in the droplet \citep{Granville2018}. 

\section{Results}
\label{sec:results}

We present our results for the HCN photochemistry as a function of the C/O ratio and UV flux (Section \ref{sec:co-uv}), and as a function of the barrier to destruction by atomic oxygen (Section \ref{sec:atomic-o}).

\subsection{The C/O Ratio, UV Flux and HCN Photochemistry}
\label{sec:co-uv}

\begin{figure}[h!]
\includegraphics[width=\textwidth]{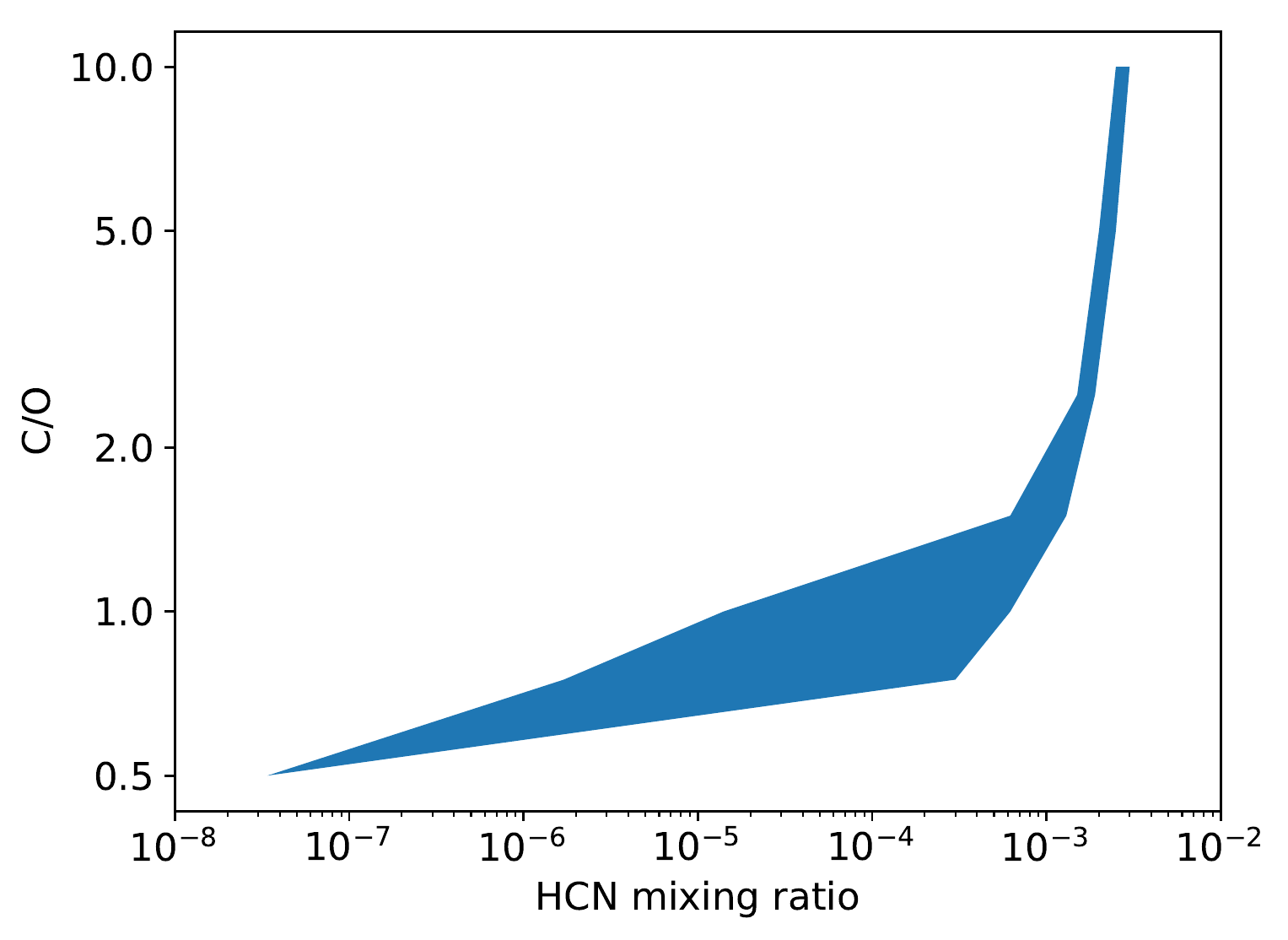}
\caption{HCN mixing ratio as a function of the C/O ratio, applied to a 0.1 mbar gas composed of various ratios of \ce{CH_4}, \ce{CO} and \ce{CO_2}, mixed with a 0.9 mbar gas composed of 99\% \ce{N_2}, 1\% \ce{H_2O} vapor and 0.1\% \ce{H_2}, exposed to the ultraviolet flux of the young Sun. The region spans several orders of magnitude, and the range can be explained by the variety of ways one can combine carbonaceous gases to achieve the same C/O ratio. For example, C/O=1 can be achieved by having a pure CO gas, or by having a gas of 50\% \ce{CO_2} and 50\% \ce{CH_4}. The composition will depend on outgassing pressure and redox state \citep{Gaillard2014}. For a given C/O ratio, the presence of methane leads to considerably more HCN.
\label{fig:c-to-o}}
\end{figure}

Regarding the C/O ratio, the photochemistry follows thermochemistry (Fig. \ref{fig:c-to-o}). Given an atmosphere dominated by \ce{N_2}:
\begin{itemize}
\item $\mathbf{\text{\bf C/O} \leq 0.5}$: The production of HCN is low, with mixing ratios of $< 1$ ppm, regardless of the particular atmospheric composition.
\item $\mathbf{0.5 < \text{\bf C/O} \leq 1.5}$: The production of HCN varies widely, with mixing ratios ranging from 1 ppm to 1000 ppm. In this regime, the particular atmospheric composition is important. If the atmosphere is 90\% \ce{N_2} and 10\% CO, HCN mixing ratios only come to 10 ppm, but if the atmosphere is 90\% \ce{N_2}, 5\% \ce{CO_2} and 5\% \ce{CH_4}, the HCN mixing ratio reaches 500 ppm. 
\item $\mathbf{\text{\bf C/O} > 1.5}$: The production of HCN is high, with mixing ratios of $> 0.1$ \% of the atmosphere, regardless of the particular atmospheric composition.
\end{itemize}

\begin{figure}[h!]
\includegraphics[width=\textwidth]{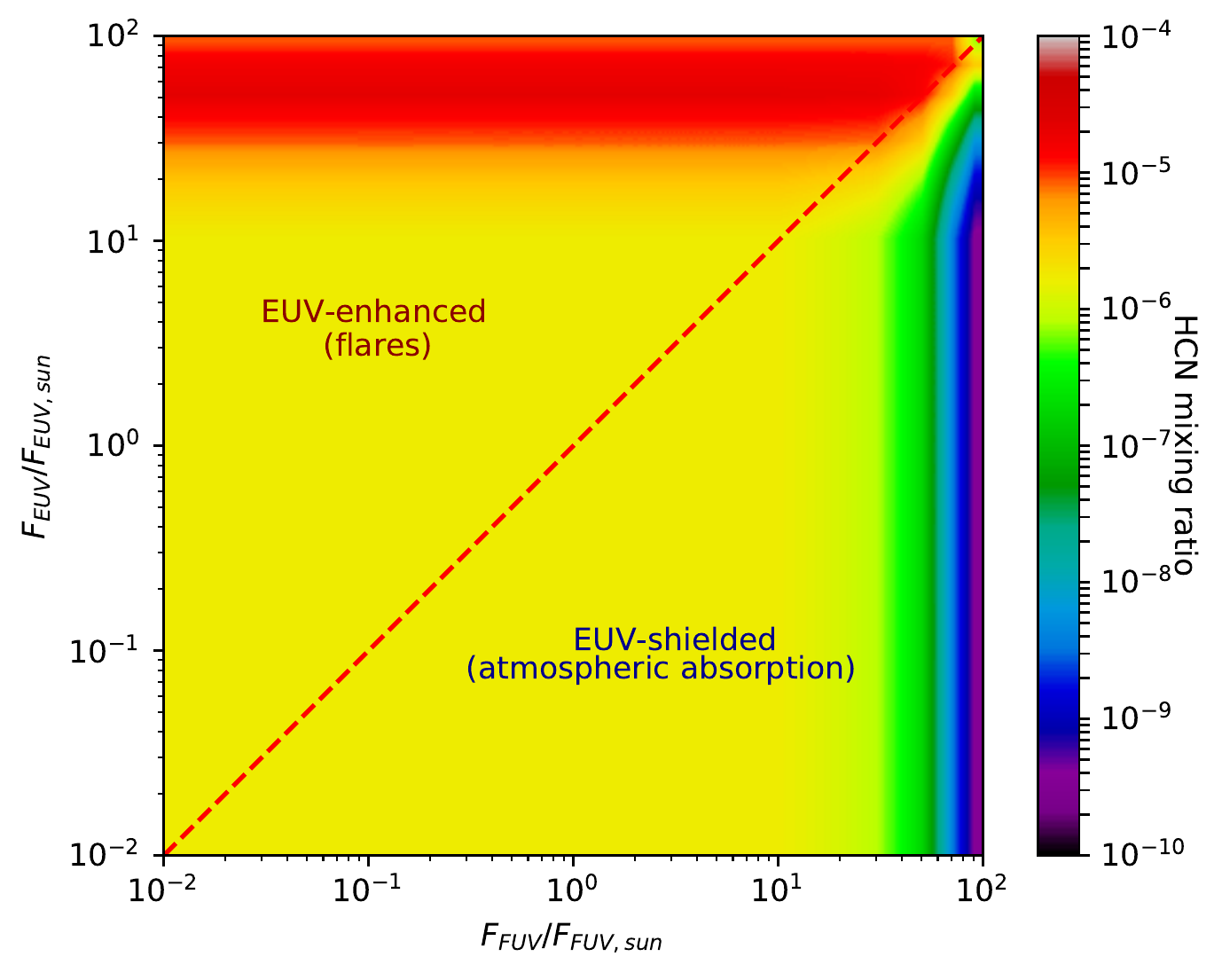}
\caption{ Ratio of a star's UV spectral irradiance to solar spectral irradiance, $F_{\lambda}/F_{\lambda,{\rm sun}}$ at EUV (500-1000~\AA) and FUV (1000-2000 \AA) wavelengths. 500-1500 \AA\ light dissociates \ce{N_2}, the first step toward forming HCN photochemically. FUV photons dissociate \ce{H_2O}, leading to hydroxyl radicals and free oxygen atoms, which rapidly destroy HCN. The dashed red line indicates where all main-sequence stellar quiescent emission lies. Stellar flares or atmospheric absorption may cause the spectrum to deviate from this line. Atmospheric absorption in particular will preferentially block the 500-1000 \AA\ photons, pushing the photoexposed atmosphere into the bottom right region of the plot. \label{fig:spec}}
\end{figure}

Photochemical production of HCN depends on the UV stellar spectrum. The EUV photons (which here we restrict to photons of wavelength 500~\AA\ -- 1000~\AA) dissociate the \ce{N_2} bond, and artificially enhancing the flux in this wavelength range (see Methods) leads to more HCN. The FUV (here between 1000~\AA\ -- 2000~\AA) flux tends to dissociate \ce{H_2O}, producing free OH and O that readily destroy HCN (see Fig. \ref{fig:pathway-to-break}). Artificially enhancing the flux in this wavelength range leads to less HCN. Fig. \ref{fig:spec} shows the effect of artificially enhancing or reducing the spectrum in these wavelength ranges. This highlights the importance of the UV environment for the HCN abundance. Stars tend to have roughly 1:1 ratios of EUV and FUV light, and so deviations that would enhance or reduce HCN will mostly be due to atmospheric attenuation at specific wavelengths of light, or spectral variability due to stellar activity (e.g. flaring).

\subsection{Atomic Oxygen as an HCN Sink and Atmospheric Methylamine}
\label{sec:atomic-o}

\begin{figure}[h!]
\includegraphics[width=\linewidth]{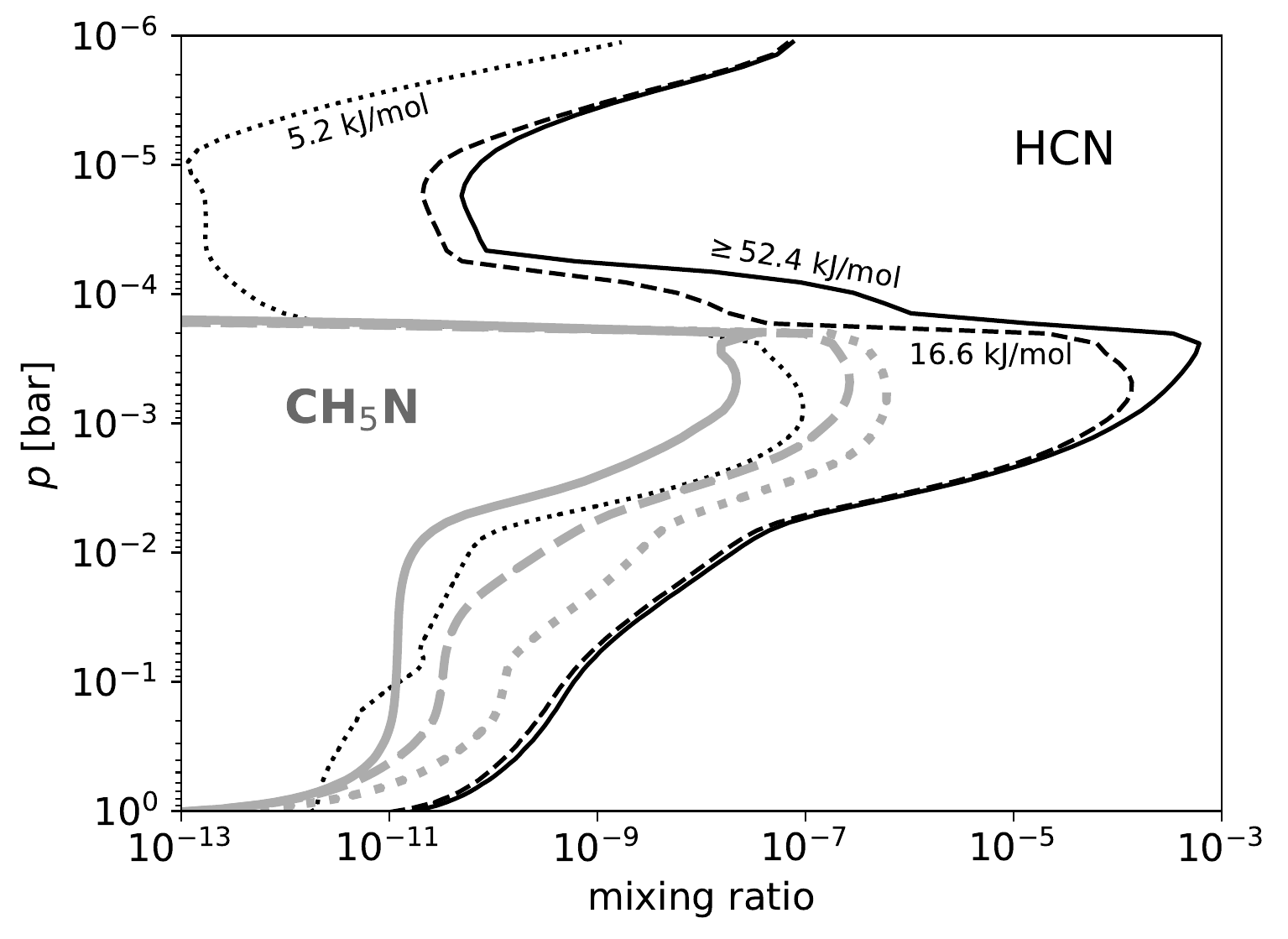}
\caption{Mixing ratios of HCN (thin black) and methylamine (\ce{CH_5N}, thick gray) as a function of pressure [bar], for a variety of energetic barriers for R\ref{eqn:HCN-dest}, ranging from 5.2 kJ/mol to 52.4 kJ/mol, with the most probable barrier lying between 16.6 and 52.4 kJ/mol, according to theory and experiment. \label{fig:hcn-react}}
\end{figure}

As with any molecule, the mixing ratio of HCN is determined by a balance of formation and loss. Hydroxyl radicals and atomic oxygen are the prime sinks of HCN. The rate constants for HCN destruction by O and OH are not all well-constrained. In particular, for the reaction:
\begin{equation}
{\rm HCN} + {\rm O} \rightarrow {\rm NCO} + {\rm H}.
\label{eqn:HCN-dest}
\end{equation}
There are several choices for the rate constant from the literature. The energy barriers, including error bars, range from $\sim 20$ kJ/mol \citep{Louge1985} to over 100 kJ/mol \citep{tsang1991}, and values in between \citep{perry1985}. We vary the rate constant of Reaction \ref{eqn:HCN-dest} for a model atmosphere of bulk composition 89\% \ce{N_2}, 10\% \ce{CO_2} and 1\% \ce{CH_4}, one of the suite of atmospheres from Rugheimer \& Rimmer \citep{Rugheimer2018}, following Rugheimer et al \citep{Rugheimer2018b}. The temperature profile is determined using a climate model, and the photochemistry is calculated with this temperature profile and a young solar evolution model (see Methods and Fig. \ref{fig:t-profile}). The results for HCN and methylamine (\ce{CH_5N}) are shown in Fig. \ref{fig:hcn-react}.

The amount of HCN in the atmosphere is clearly sensitive to the energetic barrier for R\ref{eqn:HCN-dest}. Surprisingly, the amount of methylamine in the upper atmosphere is enhanced as the energetic barrier for HCN destruction by atomic oxygen is lowered. This is because after HCN is destroyed, its products react as follows:
\begin{align}
\ce{NCO} + \ce{H} &\rightarrow \ce{CO} + \ce{NH}, \\
\ce{NH} + \ce{CH_4} + \ce{M} &\rightarrow \ce{CH_5N} + \ce{M}.
\end{align}
Near the surface ($p > 0.5$ bar), the amount of methylamine is only weakly dependent on the energy barrier for R\ref{eqn:HCN-dest}, because there is less atomic oxygen available to produce NCO. Methylamine is instead produced near the surface by:
\begin{align}
\ce{HCHO} + \ce{NO} &\rightarrow \ce{HNCO} + \ce{OH}, \\
\ce{HNCO} + h\nu &\rightarrow \ce{NH} + \ce{CO}, \\
\ce{NH} + \ce{CH_4} + \ce{M} &\rightarrow \ce{CH_5N} + \ce{M}.
\end{align}
To better assess the production of both HCN and methylamine on rocky exoplanets, the rate constants for HCN + O in particular needs to be better constrained.

\section{Discussion and Conclusions}
\label{sec:conclusion}

We have shown that \ce{HCN} abundances in the atmospheres of Earth-like rocky planets (\ce{N_2} dominated) depend critically on the atmospheric C/O ratio, with significantly greater amounts of \ce{HCN} generated photochemically when C/O $\gtrsim$ 1, regardless of the carbon-bearing species, whether it be \ce{CH_4}, \ce{CO}, or \ce{C_2H_2}. In addition, the tropospheric \ce{HCN} concentration is determined by the rate of \ce{HCN} destruction by atomic oxygen. We explore the effect of adjusting the barrier for the reaction \ce{HCN} + \ce{O} $\rightarrow$ \ce{NCO} + \ce{H}. The more efficient this reaction, the lower the tropospheric \ce{HCN} concentrations, but the higher the concentrations of methylamine (\ce{CH_5N}), a species that helps facilitate the prebiotic activation of nucleotides \citep{Mariani2018}.

When comparing our photochemistry results to alternative scenarios for forming HCN, such as impact delivery and production and production via lightning, these scenarios only require that the local C/O ratio be $\gtrsim 1$, and tend to produce \ce{HCN} in greater quantities, 0.1-1\% concentrations in the case of impacts and lightning shocks, versus 1-100 ppm tropospheric concentrations for the photochemical production of HCN. The amount of HCN generated in impacts and lightning shocks is therefore more likely to achieve concentrations of HCN used in the lab. If the global C/O ratio is $\gtrsim 1$ within the atmosphere of a rocky exoplanet, this would suggest that prebiotic chemistry starting with HCN could be initiated in any surface environment for that planet, rather than in certain more exotic scenarios. Such environmental universality may be advantageous for originating life on the surfaces of planets with long-lived reducing atmospheres.

As shown above, the formation of HCN depends on the availability of nitrogen. All of our models have started with $\sim 1$ bar of nitrogen in the form of \ce{N_2}. The concentration of \ce{N_2} will be very difficult to detect remotely on an exoplanet, owing to the molecule's lack of a dipole moment. \citet{Air2017} propose a novel observational strategy to constrain the \ce{N_2} content of exoplanet atmospheres by searching for photochemical products, especially \ce{NO}. In more reduced atmospheres, \ce{NH_3} and \ce{HCN} itself could be the major photochemical products of \ce{N_2} destruction. Either of these species would absorb at 1.5 $\mu$m, appearing as a wing on the 1.4 $\mu$m water feature, as has observed in the atmosphere of HD209458b \citep{Macdonald2017}. The shape and strength of these features will not be so strong in a temporate atmosphere with a significantly greater molecular weight. HCN will also have prominent broad features in the 7 and 14 $\mu$m regions, although at low resolution these will be difficult to disentangle from other species with similar spectra, such as \ce{C_2H_2} \citep{Tess2013}. HCN could be distinguished from these other species at high resolution. Indeed, HCN has been detected in HD209458b in emission at high resolution \citep{Hawker2018}. The search for tracers of atmospheric conditions favorable to prebiotic chemistry is beyond the scope of this paper.

The C/O ratio has been seen as an important dimension for characterizing exoplanet atmospheres\citep{Madhu2012}, and is clearly an important dimension when considering the production of hydrogen cyanide. It will likely be easier to constrain the C/O ratio for exoplanet systems in the near future than the nitrogen content, although both are essential for the photochemical production of HCN. The C/O ratio can be considered at a number of different scales, in order of decreasing size: cosmic, galactic, cluster, stellar, disk, sub-disk, planetary, surface and atmospheric C/O, and the C/O ratio within particular local environments. We will discuss each of these briefly in this order.

The cosmic C/O ratio increases over time, and within our galaxy, is expected to increase towards the galactic center, and to be enhanced in clusters with higher metallicity \citep{Esteban2005}. As low mass stars evolve and reach the asymptotic giant branch (AGB), carbon and other s-process elements are preferentially dredged up to the surface where they are released to the interstellar medium by the AGB winds \citep{Wiescher2010}. Therefore as the Universe ages, and increasingly more low mass stars evolve and die, the ratio of C/O will increase. Eventually, perhaps 30-50 Gyr from now, carbon will be more abundant than oxygen on average. This will drastically effect the geochemistry on planets produced from such an environment. 

When the C/O ratio of a star exceeds 0.8, formation models suggest that planets can be produced in the disk with overall C/O $> 1$ \citep{Bond2010}. A small fraction of stars meet this criterion \citep{Fortney2012}, and so, even today, more carbon rich areas of the disk could form these so-called ``Carbon Planets'' \citep{Kuchner2005}. When the C/O ratio is < 0.8, carbon-enhanced planets can still be formed near the midplane of the disk, sufficiently far from the host star \citep{Oberg2011}, the distance being determined by ice lines and also by the formation mechanism \citep{Booth2017}.

Once a planet's carbon is greater in abundance than oxygen by number, a few interesting chemical consequences will take place. First any oxygen will go into producing CO \citep{Lodders1997}. There will still be carbon left after the oxygen is depleted, and this will then bond with silicon, forming silicon carbide, SiC, titanium carbide, TiC, or will bind with itself to form graphite \citep{Lodders1997}. On Earth, the extra oxygen bonds with silicon forming silicates, the most abundant chemical species in Earth's crust. SiC, however, is known to be an extremely durable mineral. Since carbon bonds are stronger than oxygen, it is more difficult to melt and break these bonds. Thus plate tectonics and weathering would be more difficult with silicon carbide rocks rather than silicates. Both of these effects may cause problems in maintaining a carbon cycle on such a planet and thus a long-term habitable climate.

For stars with a lower C/O ratio, like our Sun (C/O $\approx 0.55$ \citep{Asplund2009}), formation in different parts of the disk will still affect the C/O ratio, but will probably not form ``carbon planets''. But the story doesn't end there. During and after planet formation, carbon is differentiated either into the core (carbon at high pressure becomes a siderophile \citep{Dasgupta2013}), or into the crust and atmosphere, in the form of SiC \citep{Dasgupta2013,Bond2010}, graphite \citep{Dasgupta2013,Frost2008}, or outgassed as \ce{CO_2}, CO, \ce{CH_4} \citep{Gaillard2014}, or \ce{C_2H_2} \citep{Hu2014}. At least one object in our Solar System, Titan, has an atmosphere and crust with a C/O ratio $\gg 1$ \citep{Lunine2008}. There is evidence that more than 3.8 Ga, the Earth's surface may also have had a C/O ratio $\gtrsim 1$ \citep{Yang2014}. Finally, it is worth noting that local C/O ratios vary by orders of magnitude even on the present-day Earth with its oxidized crust and atmosphere \citep{Fischer2008}, with mud volcanos releasing gas with ratios of C/O $\approx 100$ \citep{Hedberg1974,Etiope2004}. 

The debate about the C/O ratio continues \citep{Sing2016,Macdonald2017}, and hopefully the advent of JWST will be able to provide some clarity on this question. The persistent connetion between the production of HCN and the C/O ratio implies that there is a connection between a rocky planet's C/O ratio and the chances for life to originate on its surface via any chemical pathway involving HCN. This C/O ratio, along with a better understanding of stellar UV spectra \citep{Ranjan2016,Ranjan2017b,Rimmer2018c} and nitrogen content of expolanets \citep{Air2016,Air2017} may be key for determining whether a given rocky planet lies within the abiogenesis zone.

\subsection*{Acknowledgements}
P.~B.~R. acknowledges support by the Simons and Kavli Foundations for this work, including Simons Foundation SCOL awards 599634. S.~R. acknowledges support by a grant from the Simons Foundation (SCOL awards 339489). This work has made use of the MUSCLES M dwarf UV radiation field database. P.~B.~R. thanks Sukrit Ranjan and Dimitar Sasselov for helpful comments.

\bibliographystyle{elsarticle-harv}

\begin{thebibliography}{62}
\expandafter\ifx\csname natexlab\endcsname\relax\def\natexlab#1{#1}\fi
\expandafter\ifx\csname url\endcsname\relax
  \def\url#1{\texttt{#1}}\fi
\expandafter\ifx\csname urlprefix\endcsname\relax\def\urlprefix{URL }\fi

\bibitem[{Airapetian et~al.(2016)Airapetian, Glocer, Gronoff, Hebrard, and
  Danchi}]{Air2016}
Airapetian, V.~S., Glocer, A., Gronoff, G., Hebrard, E., Danchi, W., 06 2016.
  Prebiotic chemistry and atmospheric warming of early earth by an active young
  sun. Nature Geosci 9~(6), 452--455.

\bibitem[{{Airapetian} et~al.(2017){Airapetian}, {Jackman}, {Mlynczak},
  {Danchi}, and {Hunt}}]{Air2017}
{Airapetian}, V.~S., {Jackman}, C.~H., {Mlynczak}, M., {Danchi}, W., {Hunt},
  L., Nov. 2017. {Atmospheric Beacons of Life from Exoplanets Around G and K
  Stars}. Scientific Reports 7, 14141.

\bibitem[{{Ardaseva} et~al.(2017){Ardaseva}, {Rimmer}, {Waldmann}, {Rocchetto},
  {Yurchenko}, {Helling}, and {Tennyson}}]{Ardaseva2017}
{Ardaseva}, A., {Rimmer}, P.~B., {Waldmann}, I., {Rocchetto}, M., {Yurchenko},
  S.~N., {Helling}, C., {Tennyson}, J., Apr. 2017. {Lightning Chemistry on
  Earth-like Exoplanets}. ArXiv e-prints.

\bibitem[{{Asplund} et~al.(2009){Asplund}, {Grevesse}, {Sauval}, and
  {Scott}}]{Asplund2009}
{Asplund}, M., {Grevesse}, N., {Sauval}, A.~J., {Scott}, P., Sep. 2009. {The
  Chemical Composition of the Sun}. \araa 47, 481--522.

\bibitem[{Becker and Hong(1983)}]{Becker1983}
Becker, R.~S., Hong, J., 1983. Photochemistry of acetylene, hydrogen cyanide,
  and mixtures. The Journal of Physical Chemistry 87~(1), 163--166.

\bibitem[{{Bond} et~al.(2010){Bond}, {O'Brien}, and {Lauretta}}]{Bond2010}
{Bond}, J.~C., {O'Brien}, D.~P., {Lauretta}, D.~S., Jun. 2010. {The
  Compositional Diversity of Extrasolar Terrestrial Planets. I. In Situ
  Simulations}. \apj 715, 1050--1070.

\bibitem[{{Booth} et~al.(2017){Booth}, {Clarke}, {Madhusudhan}, and
  {Ilee}}]{Booth2017}
{Booth}, R.~A., {Clarke}, C.~J., {Madhusudhan}, N., {Ilee}, J.~D., Aug. 2017.
  {Chemical enrichment of giant planets and discs due to pebble drift}. \mnras
  469, 3994--4011.

\bibitem[{{Chameides} and {Walker}(1981)}]{Chameides1981}
{Chameides}, W.~L., {Walker}, J.~C.~G., Dec. 1981. {Rates of Fixation by
  Lightning of Carbon and Nitrogen in Possible Primitive Atmospheres}. Origins
  of Life 11, 291--302.

\bibitem[{{Claire} et~al.(2012){Claire}, {Sheets}, {Cohen}, {Ribas}, {Meadows},
  and {Catling}}]{claire2012}
{Claire}, M.~W., {Sheets}, J., {Cohen}, M., {Ribas}, I., {Meadows}, V.~S.,
  {Catling}, D.~C., Sep. 2012. {The Evolution of Solar Flux from 0.1 nm to 160
  {$\mu$}m: Quantitative Estimates for Planetary Studies}. \apj 757, 95.

\bibitem[{{Dasgupta}(2013)}]{Dasgupta2013}
{Dasgupta}, R., Jan. 2013. {Ingassing, Storage, and Outgassing of Terrestrial
  Carbon through Geologic Time}. Reviews in Mineralogy and Geochemistry 75,
  183--229.

\bibitem[{{Esteban} et~al.(2005){Esteban}, {Garc{\'{\i}}a-Rojas}, {Peimbert},
  {Peimbert}, {Ruiz}, {Rodr{\'{\i}}guez}, and {Carigi}}]{Esteban2005}
{Esteban}, C., {Garc{\'{\i}}a-Rojas}, J., {Peimbert}, M., {Peimbert}, A.,
  {Ruiz}, M.~T., {Rodr{\'{\i}}guez}, M., {Carigi}, L., Jan. 2005. {Carbon and
  Oxygen Galactic Gradients: Observational Values from H II Region
  Recombination Lines}. \apjl 618, L95--L98.

\bibitem[{Etiope and Milkov(2004)}]{Etiope2004}
Etiope, G., Milkov, A.~V., 2004. A new estimate of global methane flux from
  onshore and shallow submarine mud volcanoes to the atmosphere. Environmental
  Geology 46~(8), 997--1002.

\bibitem[{Ferris and Orgel(1966)}]{ferris1966}
Ferris, J.~P., Orgel, L.~E., 1966. An unusual photochemical rearrangement in
  the synthesis of adenine from hydrogen cyanide. Journal of the American
  Chemical Society 88~(5), 1074--1074.

\bibitem[{Ferus et~al.(2017)Ferus, Kubel{\'\i}k, Kn{\'\i}{\v{z}}ek, Pastorek,
  Sutherland, and Civi{\v{s}}}]{Ferus2017}
Ferus, M., Kubel{\'\i}k, P., Kn{\'\i}{\v{z}}ek, A., Pastorek, A., Sutherland,
  J., Civi{\v{s}}, S., 2017. High energy radical chemistry formation of
  hcn-rich atmospheres on early earth. Scientific reports 7~(1), 6275.

\bibitem[{{Ferus} et~al.(2015){Ferus}, {Nesvorn{\'y}}, {{\v S}poner},
  {Kubel{\'{\i}}k}, {Michal{\v c}{\'{\i}}kov{\'a}}, {Shestivsk{\'a}}, {{\v
  S}poner}, and {Civi{\v s}}}]{Ferus2015}
{Ferus}, M., {Nesvorn{\'y}}, D., {{\v S}poner}, J., {Kubel{\'{\i}}k}, P.,
  {Michal{\v c}{\'{\i}}kov{\'a}}, R., {Shestivsk{\'a}}, V., {{\v S}poner},
  J.~E., {Civi{\v s}}, S., Jan. 2015. {High-energy chemistry of formamide: A
  unified mechanism of nucleobase formation}. Proceedings of the National
  Academy of Science 112, 657--662.

\bibitem[{Fischer(2008)}]{Fischer2008}
Fischer, T.~P., 2008. Fluxes of volatiles (h2o, co2, n2, cl, f) from arc
  volcanoes. Geochemical Journal 42~(1), 21--38.

\bibitem[{{Fortney}(2012)}]{Fortney2012}
{Fortney}, J.~J., Mar. 2012. {On the Carbon-to-oxygen Ratio Measurement in
  nearby Sun-like Stars: Implications for Planet Formation and the
  Determination of Stellar Abundances}. \apjl 747, L27.

\bibitem[{{Frost} and {McCammon}(2008)}]{Frost2008}
{Frost}, D.~J., {McCammon}, C.~A., May 2008. {The Redox State of Earth's
  Mantle}. Annual Review of Earth and Planetary Sciences 36, 389--420.

\bibitem[{{Gaillard} and {Scaillet}(2014)}]{Gaillard2014}
{Gaillard}, F., {Scaillet}, B., Oct. 2014. {A theoretical framework for
  volcanic degassing chemistry in a comparative planetology perspective and
  implications for planetary atmospheres}. Earth and Planetary Science Letters
  403, 307--316.

\bibitem[{{Gao} et~al.(2015){Gao}, {Hu}, {Robinson}, {Li}, and
  {Yung}}]{Gao2015}
{Gao}, P., {Hu}, R., {Robinson}, T.~D., {Li}, C., {Yung}, Y.~L., Jun. 2015.
  {Stability of CO2 Atmospheres on Desiccated M Dwarf Exoplanets}. \apj 806,
  249.

\bibitem[{{Granville-Willett} et~al.(2018){Granville-Willett}, {Archibald},
  {Rimmer}, {Griffiths}, and {Sutherland}}]{Granville2018}
{Granville-Willett}, A., {Archibald}, A.~T., {Rimmer}, P.~B., {Griffiths}, P.,
  {Sutherland}, J.~D., 2018. {Overcoming Henry's Law}. in prep.

\bibitem[{Haqq-Misra et~al.(2008)Haqq-Misra, Domagal-Goldman, Kasting, and
  Kasting}]{haqq2008}
Haqq-Misra, J.~D., Domagal-Goldman, S.~D., Kasting, P.~J., Kasting, J.~F.,
  2008. A revised, hazy methane greenhouse for the archean earth. Astrobiology
  8~(6), 1127--1137.

\bibitem[{{Hawker} et~al.(2018){Hawker}, {Madhusudhan}, {Cabot}, and
  {Gandhi}}]{Hawker2018}
{Hawker}, G.~A., {Madhusudhan}, N., {Cabot}, S.~H.~C., {Gandhi}, S., Aug. 2018.
  {Evidence for Multiple Molecular Species in the Hot Jupiter HD 209458b}.
  \apjl 863, L11.

\bibitem[{Hedberg(1974)}]{Hedberg1974}
Hedberg, H.~D., 1974. Relation of methane generation to undercompacted shales,
  shale diapirs, and mud volcanoes. AAPG Bulletin 58~(4), 661--673.

\bibitem[{{Hu} and {Seager}(2014)}]{Hu2014}
{Hu}, R., {Seager}, S., Mar. 2014. {Photochemistry in Terrestrial Exoplanet
  Atmospheres. III. Photochemistry and Thermochemistry in Thick Atmospheres on
  Super Earths and Mini Neptunes}. \apj 784, 63.

\bibitem[{{Kharecha} et~al.(2005){Kharecha}, {Kasting}, and
  {Siefert}}]{kharecha2005}
{Kharecha}, P., {Kasting}, J.~F., {Siefert}, J., 2005. A coupled
  atmosphere--ecosystem model of the early archean earth. Geobiology 3, 53--73.

\bibitem[{{Kuchner} and {Seager}(2005)}]{Kuchner2005}
{Kuchner}, M.~J., {Seager}, S., Apr. 2005. {Extrasolar Carbon Planets}. ArXiv
  Astrophysics e-prints.

\bibitem[{Lodders and Fegley~Jr(1997)}]{Lodders1997}
Lodders, K., Fegley~Jr, B., 1997. In astrophysical implications of the
  laboratory study of presolar materials; bernatowicz, tj; zinner, e., eds. In:
  AIP Conference Proceedings. Vol. 402. p. 391.

\bibitem[{Louge and Hanson(1985)}]{Louge1985}
Louge, M.~Y., Hanson, R.~K., 1985. Shock tube study of nco kinetics. In:
  Symposium (International) on Combustion. Vol.~20. Elsevier, pp. 665--672.

\bibitem[{{Lunine} and {Atreya}(2008)}]{Lunine2008}
{Lunine}, J.~I., {Atreya}, S.~K., Mar. 2008. {The methane cycle on Titan}.
  Nature Geoscience 1, 159--164.

\bibitem[{{MacDonald} and {Madhusudhan}(2017)}]{Macdonald2017}
{MacDonald}, R.~J., {Madhusudhan}, N., Aug. 2017. {HD 209458b in new light:
  evidence of nitrogen chemistry, patchy clouds and sub-solar water}. \mnras
  469, 1979--1996.

\bibitem[{{Madhusudhan}(2012)}]{Madhu2012}
{Madhusudhan}, N., Oct. 2012. {C/O Ratio as a Dimension for Characterizing
  Exoplanetary Atmospheres}. \apj 758, 36.

\bibitem[{Mariani et~al.(2018)Mariani, Russell, Javelle, and
  Sutherland}]{Mariani2018}
Mariani, A., Russell, D.~A., Javelle, T., Sutherland, J.~D., 2018. A
  light-releasable potentially prebiotic nucleotide activating agent. Journal
  of the American Chemical Society 140~(28), 8657--8661.

\bibitem[{Miller(1957)}]{Miller1957}
Miller, S.~L., 1957. The mechanism of synthesis of amino acids by electric
  discharges. Biochimica et Biophysica Acta 23, 480--489.

\bibitem[{{{\"O}berg} et~al.(2011){{\"O}berg}, {Murray-Clay}, and
  {Bergin}}]{Oberg2011}
{{\"O}berg}, K.~I., {Murray-Clay}, R., {Bergin}, E.~A., Dec. 2011. {The Effects
  of Snowlines on C/O in Planetary Atmospheres}. \apjl 743, L16.

\bibitem[{{Patel} et~al.(2015){Patel}, {Percivalle}, {Ritson}, {Duffy}, and
  {Sutherland}}]{patel2015}
{Patel}, B.~H., {Percivalle}, C., {Ritson}, D.~J., {Duffy}, C.~D.,
  {Sutherland}, J.~D., Apr. 2015. {Common origins of RNA, protein and lipid
  precursors in a cyanosulfidic protometabolism}. Nature Chemistry 7, 301--307.

\bibitem[{Pavlov et~al.(2000)Pavlov, Kasting, Brown, Rages, and
  Freedman}]{pavlov2000}
Pavlov, A.~A., Kasting, J.~F., Brown, L.~L., Rages, K.~A., Freedman, R., 2000.
  Greenhouse warming by ch4 in the atmosphere of early earth. Journal of
  Geophysical Research: Planets 105~(E5), 11981--11990.

\bibitem[{Perry and Melius(1985)}]{perry1985}
Perry, R.~A., Melius, C.~F., 1985. The rate and mechanism of the reaction of
  hcn with oxygen atoms over the temperature range 540--900 k. In: Symposium
  (International) on Combustion. Vol.~20. Elsevier, pp. 639--646.

\bibitem[{{Pierrehumbert} and {Gaidos}(2011)}]{Wordsworth2011}
{Pierrehumbert}, R., {Gaidos}, E., Jun. 2011. {Hydrogen Greenhouse Planets
  Beyond the Habitable Zone}. \apjl 734, L13.

\bibitem[{{Ranjan} and {Sasselov}(2016)}]{Ranjan2016}
{Ranjan}, S., {Sasselov}, D.~D., Jan. 2016. {Influence of the UV Environment on
  the Synthesis of Prebiotic Molecules}. Astrobiology 16, 68--88.

\bibitem[{{Ranjan} and {Sasselov}(2017)}]{Ranjan2017b}
{Ranjan}, S., {Sasselov}, D.~D., Mar. 2017. {Constraints on the Early
  Terrestrial Surface UV Environment Relevant to Prebiotic Chemistry}.
  Astrobiology 17, 169--204.

\bibitem[{{Rimmer} and {Helling}(2016)}]{Rimmer2016}
{Rimmer}, P.~B., {Helling}, C., May 2016. {A Chemical Kinetics Network for
  Lightning and Life in Planetary Atmospheres}. The Astrophysical Journal
  Supplement Series 224, 9.

\bibitem[{Rimmer et~al.(2018)Rimmer, Xu, Thompson, Gillen, Sutherland, and
  Queloz}]{Rimmer2018c}
Rimmer, P.~B., Xu, J., Thompson, S.~J., Gillen, E., Sutherland, J.~D., Queloz,
  D., 2018. The origin of rna precursors on exoplanets. Science Advances 4~(8),
  eaar3302.

\bibitem[{Ritson and Sutherland(2012)}]{ritson2012}
Ritson, D., Sutherland, J.~D., 2012. Prebiotic synthesis of simple sugars by
  photoredox systems chemistry. Nature chemistry 4~(11), 895.

\bibitem[{{Rugheimer} and {Kaltenegger}(2018)}]{Rugheimer2018b}
{Rugheimer}, S., {Kaltenegger}, L., Feb. 2018. {Spectra of Earth-like Planets
  through Geological Evolution around FGKM Stars}. \apj 854, 19.

\bibitem[{{Rugheimer} and {Rimmer}(2018)}]{Rugheimer2018}
{Rugheimer}, S., {Rimmer}, P.~B., 2018. {Exoplanet Prebiosignatures}. in prep.

\bibitem[{Saladino et~al.(2001)Saladino, Crestini, Costanzo, Negri, and
  Di~Mauro}]{Saladino2001}
Saladino, R., Crestini, C., Costanzo, G., Negri, R., Di~Mauro, E., 2001. A
  possible prebiotic synthesis of purine, adenine, cytosine, and 4
  (3h)-pyrimidinone from formamide: implications for the origin of life.
  Bioorganic \& medicinal chemistry 9~(5), 1249--1253.

\bibitem[{Sanchez et~al.(1967)Sanchez, Ferbis, and Orgel}]{Sanchez1967}
Sanchez, R.~A., Ferbis, J.~P., Orgel, L.~E., 1967. Studies in prebiodc
  synthesis: Ii. synthesis of purine precursors and amino acids from aqueous
  hydrogen cyanide. Journal of molecular biology 30~(2), 223--253.

\bibitem[{Sanchez et~al.(1966)Sanchez, Ferris, and Orgel}]{sanchez1966}
Sanchez, R.~A., Ferris, J.~P., Orgel, L.~E., 1966. Cyanoacetylene in prebiotic
  synthesis. Science 154~(3750), 784--785.

\bibitem[{Sanchez and Orgel(1970)}]{sanchez1970}
Sanchez, R.~A., Orgel, L.~E., 1970. Studies in prebiotic synthesis: V.
  synthesis and photoanomerization of pyrimidine nucleosides. Journal of
  molecular biology 47~(3), 531--543.

\bibitem[{{Sing} et~al.(2016){Sing}, {Fortney}, {Nikolov}, {Wakeford},
  {Kataria}, {Evans}, {Aigrain}, {Ballester}, {Burrows}, {Deming},
  {D{\'e}sert}, {Gibson}, {Henry}, {Huitson}, {Knutson}, {Lecavelier Des
  Etangs}, {Pont}, {Showman}, {Vidal-Madjar}, {Williamson}, and
  {Wilson}}]{Sing2016}
{Sing}, D.~K., {Fortney}, J.~J., {Nikolov}, N., {Wakeford}, H.~R., {Kataria},
  T., {Evans}, T.~M., {Aigrain}, S., {Ballester}, G.~E., {Burrows}, A.~S.,
  {Deming}, D., {D{\'e}sert}, J.-M., {Gibson}, N.~P., {Henry}, G.~W.,
  {Huitson}, C.~M., {Knutson}, H.~A., {Lecavelier Des Etangs}, A., {Pont}, F.,
  {Showman}, A.~P., {Vidal-Madjar}, A., {Williamson}, M.~H., {Wilson}, P.~A.,
  Jan. 2016. {A continuum from clear to cloudy hot-Jupiter exoplanets without
  primordial water depletion}. \nat 529, 59--62.

\bibitem[{Strecker(1854)}]{Strecker1854}
Strecker, A., 1854. Ueber einen neuen aus aldehyd-ammoniak und blaus{\"a}ure
  entstehenden k{\"o}rper. Justus Liebigs Annalen der Chemie 91~(3), 349--351.

\bibitem[{{Tessenyi} et~al.(2013){Tessenyi}, {Tinetti}, {Savini}, and
  {Pascale}}]{Tess2013}
{Tessenyi}, M., {Tinetti}, G., {Savini}, G., {Pascale}, E., Sep. 2013.
  {Molecular Detectability in Exoplanetary Emission Spectra}. European
  Planetary Science Congress 8, EPSC2013--817.

\bibitem[{{Tian} et~al.(2011){Tian}, {Kasting}, and {Zahnle}}]{Tian2011}
{Tian}, F., {Kasting}, J.~F., {Zahnle}, K., Aug. 2011. {Revisiting HCN
  formation in Earth's early atmosphere}. Earth and Planetary Science Letters
  308, 417--423.

\bibitem[{{Toon} et~al.(1989){Toon}, {McKay}, {Ackerman}, and
  {Santhanam}}]{Toon1989}
{Toon}, O.~B., {McKay}, C.~P., {Ackerman}, T.~P., {Santhanam}, K., Nov. 1989.
  {Rapid calculation of radiative heating rates and photodissociation rates in
  inhomogeneous multiple scattering atmospheres}. Journal of Geophysical
  Research: Atmospheres 94, 16287--16301.

\bibitem[{{Tsai} et~al.(2017){Tsai}, {Lyons}, {Grosheintz}, {Rimmer},
  {Kitzmann}, and {Heng}}]{Tsai2017}
{Tsai}, S.-M., {Lyons}, J.~R., {Grosheintz}, L., {Rimmer}, P.~B., {Kitzmann},
  D., {Heng}, K., Feb. 2017. {VULCAN: An Open-source, Validated Chemical
  Kinetics Python Code for Exoplanetary Atmospheres}. \apjs 228, 20.

\bibitem[{Tsang and Herron(1991)}]{tsang1991}
Tsang, W., Herron, J.~T., 1991. Chemical kinetic data base for propellant
  combustion i. reactions involving no, no2, hno, hno2, hcn and n2o. Journal of
  physical and chemical reference data 20~(4), 609--663.

\bibitem[{{Wiescher} et~al.(2010){Wiescher}, {G{\"o}rres}, {Uberseder},
  {Imbriani}, and {Pignatari}}]{Wiescher2010}
{Wiescher}, M., {G{\"o}rres}, J., {Uberseder}, E., {Imbriani}, G., {Pignatari},
  M., Nov. 2010. {The Cold and Hot CNO Cycles}. Annual Review of Nuclear and
  Particle Science 60, 381--404.

\bibitem[{{Wordsworth} and {Pierrehumbert}(2014)}]{Wordsworth2014}
{Wordsworth}, R., {Pierrehumbert}, R., Apr. 2014. {Abiotic Oxygen-dominated
  Atmospheres on Terrestrial Habitable Zone Planets}. \apjl 785, L20.

\bibitem[{Xu et~al.(2018)Xu, Ritson, Ranjan, Todd, Sasselov, and
  Sutherland}]{Xu2018}
Xu, J., Ritson, D.~J., Ranjan, S., Todd, Z.~R., Sasselov, D.~D., Sutherland,
  J.~D., 2018. Photochemical reductive homologation of hydrogen cyanide using
  sulfite and ferrocyanide. Chemical Communications 54~(44), 5566--5569.

\bibitem[{{Yang} et~al.(2014){Yang}, {Gaillard}, and {Scaillet}}]{Yang2014}
{Yang}, X., {Gaillard}, F., {Scaillet}, B., May 2014. {A relatively reduced
  Hadean continental crust and implications for the early atmosphere and
  crustal rheology}. Earth and Planetary Science Letters 393, 210--219.

\bibitem[{{Zahnle}(1986)}]{Zahnle1986}
{Zahnle}, K.~J., Feb. 1986. {Photochemistry of methane and the formation of
  hydrocyanic acid (HCN) in the Earth's early atmosphere}. \jgr 91, 2819--2834.

\end{thebibliography}

\end{document}